\documentclass[12pt]{article}
\usepackage{latexsym,amsfonts,amssymb,mathrsfs,amsmath,amsthm,caption,subfigure,cite}
\usepackage[pagewise]{lineno}
\usepackage[dvips]{color}
\usepackage{colordvi,multicol}
\usepackage{graphicx}
\usepackage{bm}
\usepackage{CJK}
\usepackage{multirow}
\usepackage{float}
\usepackage{bbding}

\newtheorem{thm}{Theorem}[section]
\newtheorem{cor}[thm]{Corollary}

\newtheorem{rem}[thm]{Remark}

\makeatletter
\def\@makefnmark{}
\makeatother

\begin{document}
\title{\bf {Optimal Control of Investment for an  Insurer
 in Two Currency Markets}}

\renewcommand{\thefootnote}{}
\footnotetext{ \noindent{This work is supported by the National Natural Science Foundation of China (Grant No. 11931018) and Tianjin Natural Science Foundation.} \\
$^{+}$ Corresponding author.
}
\author{\bf   {Qianqian Zhou,      Junyi Guo$^{+}$}\\
\small { {School of Mathematical Sciences,
 Nankai University,  Tianjin,  China}}\\
 \small{(Email: qianqzhou@yeah.net, jyguo@nankai.edu.cn)}}

\date{}
\maketitle

\vskip 0.5cm \noindent{\bf Abstract}\quad
In this paper, we study the optimal investment problem of an insurer whose   surplus process  follows  the diffusion approximation of the classical Cramer-Lundberg model. Investment in  the foreign   market is allowed, and therefore, the foreign exchange rate  model is considered and incorporated. It is assumed that the instantaneous mean growth rate of foreign exchange rate price  follows an Ornstein-Uhlenbeck process.  Dynamic programming method is employed to study the problem of maximizing the expected exponential utility of terminal wealth. By solving the corresponding Hamilton-Jacobi-Bellman equations,  the optimal investment strategies and the value functions are obtained. Finally, numerical analysis is presented.

\smallskip

\noindent {\bf Keywords:}\quad  Cramer-Lundberg model; Exponential utility; Hamilton-Jacobi-Bellman equation; Optimal investment strategy;   Foreign exchange rate

\smallskip

\noindent {\bf 2000 MR Subject Classification}\quad  62P05; 91B30; 93E20

\section{Introduction}
In actuarial science and applied probability,  risk  theory  is a traditional and modern field, which uses mathematical models to describe an insurer's vulnerability to ruin.
In order to decrease (increase) the risk (profit), the insurance companies are allowed to  invest their wealth into risk-free assets  and risky assets. And in recent years,  there are many remarkable works of   optimal investment problems. Especially, maximizing the expected exponential utility function  and  minimizing the probability of ruin have attracted a substantial amount of interest.

Browne \cite{B95} used Brownian motion with drift to describe the surplus of the insurance
company and found the optimal investment strategy to maximize  the expected exponential
utility of the terminal wealth and minimize  the probability of ruin. Later, Yang and Zhang \cite{YZ05} explored the same optimal investment problem for a risk process modeled by a jump diffusion process. In Hipp and Pulm \cite{HP00}, the authors considered a risk process modeled as a compound Poisson process  and investigated the optimal investment strategy to minimize the ruin probability of this model. Liu and Yang \cite{LY04} generalized the model  in  Hipp and Pulm \cite{HP00}  by including a risk-free asset. And then  the optimal investment strategy was investigated. Schmidli \cite{SH02}  also studied  the compound Poisson risk model  and the optimal investment strategy of minimizing its ruin probability. In Bai and Guo \cite{BG08}, they considered the optimal problem with multiple risky assets  and no-shorting constraint.
By solving the corresponding Hamilton-Jacobi-Bellman equations, the optimal strategies for maximizing the expected exponential utility and minimizing the ruin probability were obtained.
Wang \cite{WN07} considered the optimal investment strategy to maximize the exponential utility of an insurance company's reserve. The claim process was supposed to be a pure jump process, which is  not necessarily compound Poisson, and the insurer has the option of investing in multiple risky assets.

However, prior studies did not  consider the condition that the insurers are allowed
to invest their wealth in more than one currency market. Thus, in this paper  we investigate the case that the insurance company is allowed to invest its wealth into more than one currency market, such that it can invest its  wealth  into domestic risk-free  assets   and foreign   risky assets.

The connection between the domestic currency market and the foreign currency market is the  exchange rate.  Foreign exchange rate  plays an important role as a tool used to convert foreign market cash flows into domestic currency.  And there are many factors that influence exchange rate prices, such as inflation, balance of international payment, interest-rate spread,   etc.

Inflation is the most important fundamental factor affecting the movements of exchange rate. If the inflation rate of a domestic country is higher than that of a foreign country, the competitiveness of the domestic country's exports is weakened, while this increases  the competitiveness of foreign goods in  domestic country's market.  It would cause the domestic country's trade balance of payments deficit and the demand of foreign exchange is larger than the supply. Then it leads to the increase of  foreign exchange rate price. Conversely, the price of  foreign  exchange rate  declines.

The balance of payments is the  direct factor that affects the exchange rate. For example, when a  country has a large balance of payments surplus, i.e., the country's imports are less than its exports, its currency demand will increase, which will lead to an increase in foreign exchange flowing into the country. In this way, in the foreign exchange market, the supply of foreign exchange is greater than the demand then the foreign exchange rate price goes down. But if  a country has a large balance of payments deficit, i.e., the country's imports are more than its exports, the supply of foreign exchange is greater than the demand which  leads to the increase of foreign exchange  rate price.

Under certain conditions, interest rates have a great short-term impact on  exchange rate price. This effect  is caused by the difference of interest rates between  different countries.
In general, if interest- rate spread  is increasing then the demand for  domestic  currency  increases. This  leads to the increase of the domestic currency price. And thus the foreign exchange rate price declines.   Conversely, if the interest- rate spread   goes down, the price of foreign exchange rate is increasing.

As we all known, the classical  geometric Brownian motion is  the most commonly used model to describe the dynamics  of  exchange rate price. In Musiela and Rutkowski \cite{MMR05}, the explicit valuation formulas for various kinds of currency and foreign equity options were established in which the foreign exchange rate was modeled by means of geometric Brownian motion.
Veraart \cite{V11}  considered an investor in the foreign exchange market who trades  in  domestic  currency market and foreign currency market  with the exchange rate modeled as a geometric Brownian motion. After that, a more general model of exchange rate was used. In Eisenberg \cite{E16}, the author considered an insurance company seeking to maximize the expected discounted dividends whereas the dividends are declared or paid in a foreign currency. It was assumed that the insurance company generates  its income in a foreign currency but pays dividends in its home currency. In that paper, the exchange rate was modeled by a geometric L$\acute{e}$vy process.  However, taking  into account of  a  variety of  factors which affect the  exchange rate, the classical  geometric Brownian motion can not  better reflect the real dynamics  of  exchange rate. Thus other models have been created to describe  exchange rate. One of the most popular model  is the one in  which  the interest- rate spread is   incorporated  into the geometric Brownian motion. For example,
 in Guo, et al. \cite{GZP18}, the authors investigated  the optimal  strategy of an insurer who invests  in both domestic and foreign markets. They assumed that  the domestic and  foreign
 nominal interest rates are both described by extended Cox-Ingersoll-Ross (CIR)  model. And the exchange rate price  is modeled  by  geometric Brownian motion with  domestic and foreign interest rates.
  For more details of  the model of  exchange rate price described by    geometric Brownian motion  with interest rates one can refer    \cite{RA08, RM13, RM16}.

Although interest-rate spread has a certain impact on  exchange rate price and  the models  with interest-rate spread  are excellent,   from the perspective of the basic factors determining the trend of exchange rate fluctuation  that the effect of interest- rate spread is limited. Moreover, from the above descriptions we obtain that the supply and demand of domestic and foreign currencies is the most paramount and direct factor on the change of exchange rate price.
Thus in our paper,  the foreign exchange market's supply and demand  is incorporated into  the model of exchange rate price.

In Liang, et al. \cite{LYG11}   and Rishel \cite{RR99}, they  used the  model of geometric Brownian motion, where the mean growth rate is given by  Ornstein-Uhlenbeck process,  to describe the dynamics of risky asset price which can have features of bull and bear markets. Inspired by the models of risky assets in  Liang, et al. \cite{LYG11}   and Rishel \cite{RR99}  and the discussions of the effect factors of foreign exchange rate, we consider that the foreign exchange rate also has the bull and bear markets. That is
when the demand of foreign currency is larger than the supply,  the foreign exchange rate price increases, we call it   the "bull foreign exchange  rate". In addition,  when the demand of foreign currency is less than the supply, the foreign exchange rate price goes down, we call it   the "bear foreign exchange  rate". The commonly-used model for the exchange rate price is the geometric Brownian motion in which the expected instantaneous rate and the volatility of the exchange rate price are both constants. This seems to rule out bull and bear markets.

In this paper,  the price of foreign  exchange rate $Q_t$ is described by the following differential equation:
\begin{gather}\label{a2}
dQ_t=Q_t\big\{a(t)dt +\sigma_{Q}dW^2_t\big\},\     Q_0=q,
\end{gather}
where $a(t)=u_Q+m(t)$  and  $m(t)$  is given by the Ornstein-Uhlenbeck   equation
$$dm(t)=\alpha m(t)dt+\beta dW^3_t,\      m(0)=m_0.$$
Here $ u_Q, \sigma_Q, q, \alpha, \beta$  are known constants and they are all positive except $\alpha$  and $\beta.$ In (\ref{a2}), $u_Q$  is the target  mean growth rate for the  exchange rate. If $m(t)>0,$  then $a(t)$  is substantially larger than $u_Q,$  this could be considered as a bull market of exchange rate. Conversely, if $m(t)<0$ then $a(t)$  is
substantially  less than $u_Q,$ this  could be considered as a bear market of exchange rate. Especially, if $a(t)<0$   the  exchange rate price  goes down. Here the function of $m(t)$   is to let the random mean growth rate of foreign exchange rate price  be close to the target  mean growth rate. Once the mean growth rate of foreign exchange rate price is larger than the target  mean growth rate for a long time, then $m(t)<0.$ Otherwise, $m(t)>0.$

The insurance company is allowed to invest its wealth into domestic and foreign currency markets with the exchange rate price  $Q_t$ described by (\ref{a2}).  Our target is to maximize the expected exponential utility of terminal wealth over all admissible strategies.

The rest of the paper is organized as follows. The model is described in Section \ref{sec2}. Our main results are given in Section \ref{sec3}. By solving the corresponding Hamilton-Jacobi-Bellman equations, the optimal value functions and optimal strategies  are explicitly derived. In particular, we find that if the insurance company only invests in foreign risky assets and the price of exchange rate is modeled by geometric Brownian motion then  the optimal investment strategy is a constant, regardless  of the level of wealth the company has. In  the last section,  numerical examples and analysis are presented. And we find that, in some cases, investing   into  two currency markets  can produce a higher value function than investing into only one currency market.

\section{The model} \label{sec2}
We start from  the classical Cramer-Lundberg model in which the  surplus of the insurance company is modeled as
$$X_{t}=X_{0}+pt-\sum_{i=1}^{N(t)}Z_{i},\text{ with } X_0=x,$$
where $p$ is the premium rate, $N(t)$ is the Poisson process stating the number of claims,  and $Z_{i}$ is a sequence of independent  random variables which are identically distributed representing the size of  claims. Without the loss  of generality we assume  that the intensity of the process $N(t)$  is 1, then the  dynamics of $X_t$  can be approximated by
$$dX_{t}=udt+\sigma dW_{t},\  \  X_0=x,$$
where $u=p-E[Z]>0, \sigma^{2}=E[Z^{2}],$   and $W_{t}$ is a standard Brownian motion.
 For more details of the  diffusion approximation of the surplus process, we can refer to \cite{EH75, G89, IDL69, SH94}.

The insurance company invests its wealth into domestic risk-free asset and foreign risky asset.    The price of domestic risk-free asset is given by
\begin{gather*}
dB_{t}^{d}=B_{t}^{d}r_{d}dt,\  \  B^d(0)=B^d_0.
\end{gather*}
The foreign risky asset price $S^{f}_t$ is modeled by means of geometric Brownian motion such  that
\begin{gather}\label{a1}
dS_{t}^{f}=S_{t}^{f}(u_{f}dt+\sigma_{f}dW_{t}^{1}),\  \  S^f(0)=S^f_0.
\end{gather}

We adopt here the convention that the price $S_{t}^{f}$ is denominated by foreign currency and the exchange rate is denominated in units of domestic currency per unit of foreign currency.
This means that $Q_{t}$ represents the domestic price at time $t$ of one unit of the foreign currency.
Thus let  $g_{t}:=g(S_{t}^{f}, Q_{t})=Q_{t}S_{t}^{f},$ then $g_{t}$ is the price of foreign risky asset denominated by domestic currency. By It$\hat{o}^{'}$s formula and the formulas of (\ref{a2}) and (\ref{a1})   we find  that   $g_{t}$ satisfies the following stochastic differential equation
$$dg_{t}=g_{t}\big\{(u_{f}+a(t))dt+\sigma_{f}dW_{t}^{1}+\sigma_{Q}dW_{t}^{2}\big\}.$$

The  total amount of money invested in  foreign risky asset at time $t$ is denoted by $\pi_{t}$  and  the rest of the surplus is invested into domestic risk-free asset.
Under the strategy $\pi_t,$  the surplus of the insurance company is as follows:
\begin{eqnarray}\label{a3}
d X_{t}^{\pi}=\big\{ \pi_t A_1+u+r_d X_t+\pi_t m(t) \big\}dt+\sigma dW_t+\pi_t \sigma_{f}d W^{1}_t+\pi_t \sigma_{Q}d W^{2}_t,
\end{eqnarray}
where $A_1=u_f+u_{Q}-r_d$  and  the initial surplus  is $X_0^{\pi}=x_0.$
 Here it is allowed  that  $\pi_t<0$  and $\pi_t>X^{\pi}_t$  which  means  that the company is allowed to short sell  the foreign risky asset and borrow money for  investment in foreign risky asset. And
we assume  that
 $W_{t}, W_{t}^{1}, W^{2}_{t}, W^3_t$ are independent standard  Brownian motions on the same probability space $(\Omega, \mathcal F, P).$

  We are going to maximize the expected exponential utility of terminal wealth over all admissible strategies $\pi_t.$ A strategy $\pi_{t}$ is said to be admissible, if $\pi_{\cdot}$  is $\mathcal{F}_t$-adapted, where $\mathcal{F}_t$   is the filtration generated by $X^{\pi}_t,$ and  for any $T>0,$ $E[\int_{0}^{T}\pi^{2}(t) dt]<\infty.$ The set of all admissible strategies is denoted by $\Pi.$

\section{The main results}\label{sec3}
Suppose now that the insurance company  is interested in maximizing the utility of  its  terminal wealth at time $T.$ Denote the  utility function as  $u(x)$ with $u^{'}(x)>0$ and $u^{''}(x)<0.$ For a strategy $\pi,$  the utility attained by the insurer from state $x, m$ at time $t$ is defined as
$$V_{\pi}(t,x,m)=E[u(X_{T}^{\pi})|(X_{t}^{\pi}, m(t))=(x,m)].$$
Our objective is to find the optimal value function
\begin{gather}\label{a0}
V(t,x,m)=\sup_{\pi\in \Pi}V_{\pi}(t,x,m)
\end{gather}
and the optimal investment strategy $\pi^{*}$ such that $V_{\pi^{*}}(t,x,m)=V(t,x,m).$

Assume now that the investor has an exponential utility function
\begin{eqnarray}\label{2}
u(x)=\lambda-\frac{\gamma}{\theta}e^{-\theta x},
\end{eqnarray}
where $\gamma>0$ and $\theta>0.$ The utility function  (\ref{2}) plays a remarkable part in insurance mathematics and actuarial practice, since it is the only utility function under which the principle of "zero utility" gives a fair premium that is independent of the level of reserve of an insurance company (see Gerber \cite{GH79}).

 Applying   the dynamic programming approach described in \cite{FR93}, from standard arguments, we see that if the optimal value function $V(t,x,m)$ and its partial derivatives $V_{t}, V_{x}, V_{xx}, V_{m}, V_{mm}$ are continuous on $[0,T]\times R^{1}\times R^{1},$ then $V(t,x,m)$ satisfies the following Hamilton-Jacobi-Bellman (HJB) equation
\begin{eqnarray}\label{3}
V_t&+&\sup_{\pi}\big\{[\pi A_1+\pi m+xr_d+u] V_x+\frac{1}{2}[\sigma^2+\pi^2 (\sigma^2_f+\sigma^2_Q)]V_{xx}     \big\}\nonumber\\
&+&\alpha m V_m+\frac{1}{2}\beta^2 V_{mm}=0,
\end{eqnarray}
with boundary condition
$V(T,x,m)=u(x).$

In order to solve the HJB equation (\ref{3}), we first find the value $\pi(x,m)$  which maximizes the function
\begin{gather}\label{a3+}
\big(\pi A_1+\pi m+xr_d+u\big) V_x+\frac{1}{2}[\sigma^2+\pi^2 (\sigma^2_f+\sigma^2_Q)]V_{xx}
\end{gather}
Differentiating with respect to $\pi$ in (\ref{a3+}) the optimizer
\begin{gather}\label{a3++}
\pi^{*}=-\frac{A_1+m}{\sigma^2_f+\sigma^2_Q}\frac{V_x}{V_{xx}}
\end{gather}
is obtained.

Assume  that HJB equation (\ref{3})  has a  classical solution $V$ such that
$V_x > 0$ and $V_{xx} < 0.$
Inspired by the form of the solution in  \cite{B95}, we try to find the  solution of   (\ref{3})  as the form
\begin{gather}\label{a4}
V(t,x,m)=\lambda-\frac{\gamma}{\theta}exp\big\{ -\theta x e^{r_{d}(T-t)}+h(t,m)   \big\},
\end{gather}
where $h(t,m)$  is a suitable function such that (\ref{a4}) is a solution  of (\ref{3}). And the boundary  condition $V(T,x,m)=u(x)$  implies  that $h(T,m)=0.$

From (\ref{a4})  we  can calculate that
\begin{gather*}
V_t=\big[ V(t,x,m)-\lambda  \big] \big\{ \theta x r_d e^{r_d(T-t)}+h_t  \big\}\\
V_x=-\big[ V(t,x,m)-\lambda  \big]\theta e^{r_d(T-t)},\  \  V_{xx}=\big[ V(t,x,m)-\lambda  \big]\theta^2 e^{2r_d(T-t)}\\
V_m=\big[ V(t,x,m)-\lambda  \big]h_m,\  \  V_{mm}=\big[ V(t,x,m)-\lambda  \big](h^2_m+h_{mm}),
\end{gather*}
where  $V_t, V_{x}, V_{xx}, V_{m}, V_{mm}$  are the partial derivatives of $V(t,x,m)$  and $h_t, h_m, h_{mm}$  are the partial derivatives of $h(t,m).$  Substituting $V_t, V_x, V_{xx}, V_m, V_{mm}$   back into (\ref{3}) yields
\begin{eqnarray}\label{a5}
h_t&+&\sup_{\pi}\Big\{-\pi(A_1+m)\theta e^{r_d(T-t)}-u \theta e^{r_d(T-t)}+\frac{1}{2}\pi^2\theta^2(\sigma^2_f+\sigma^2_Q)e^{2r_d(T-t)}    \Big\}\nonumber\\
&+&\frac{1}{2}\theta^2\sigma^2e^{2r_d(T-t)}+\alpha m h_m+\frac{1}{2}\beta^2(h^2_m+h_{mm})=0.
\end{eqnarray}
And from (\ref{a3++})
\begin{gather}\label{a6}
\pi^*=\frac{A_1+m}{\theta(\sigma^2_f+\sigma^2_{Q})}e^{-r_d(T-t)}.
\end{gather}
Put    $\pi^{*}$  into  (\ref{a5})  and  calculate   then
\begin{eqnarray}\label{a7}
h_t-u\theta e^{r_d(T-t)}+\frac{1}{2}\theta^2 \sigma^2 e^{2r_d(T-t)}-\frac{1}{2}\frac{(A_1+m)^2}{\sigma^2_f+\sigma^2_{Q}}
+\alpha mh_m+\frac{1}{2} \beta^2(h_m^2+h_{mm})=0.
\end{eqnarray}
It can be shown that  (\ref{a4}) is a solution to  (\ref{a5}) if  $h(t,m)$ is a solution to (\ref{a7}).

\begin{thm}\label{thm1}
With the terminal condition $h(T,m)=0,$    the partial differential equation (\ref{a7})  has the solution of the form
\begin{gather}\label{thm1-a}
h(t,m)=K(t)m^2+L(t)m+J(t),
\end{gather}
where $K(t)$  is a solution to
\begin{gather}\label{thm1-b}
K^{'}(t)+2\beta^2K^2(t)+2\alpha K(t)-\frac{1}{2(\sigma^2_f+\sigma^2_Q)}=0,\   \   K(T)=0;
\end{gather}
L(t) is a solution to
\begin{gather}\label{thm1-c}
L^{'}(t)+(\alpha+2\beta^2 K(t))L(t)-\frac{A_1}{\sigma^2_f+\sigma^2_Q}=0,\  \  L(T)=0;
\end{gather}
and J(t)  is a solution to
\begin{gather}\label{thm1-d}
J^{'}(t)-u\theta e^{r_d(T-t)}+\frac{1}{2}\theta^2 \sigma^2 e^{2r_d(T-t)}-\frac{A^2_1}{2(\sigma^2_f+\sigma^2_Q)}+\frac{1}{2}\beta^2 L^2+\beta^2 K=0,\  \ J(T)=0.
\end{gather}

\end{thm}
{\bf Proof. }  Substituting  (\ref{thm1-a})  into  (\ref{a7})  and combining like terms with respect to the powers of $m,$  we have that
\begin{eqnarray}
&&m^2\big\{ K^{'}(t)+2\beta^2K^2(t)+2\alpha K(t)-\frac{1}{2(\sigma^2_f+\sigma^2_Q)}   \big\}+\nonumber\\
&&m\big\{L^{'}(t)+\alpha L(t)+2\beta^2 K(t)L(t)-\frac{A_1}{\sigma^2_f+\sigma^2_Q}   \big\}+
\big\{ J^{'}(t)\nonumber\\
&&-u\theta e^{r_d(T-t)}+\frac{1}{2} \theta^2\sigma^2 e^{2r_d(T-t)}-\frac{A^2_1}{2(\sigma^2_f+\sigma^2_Q)} +\frac{1}{2}\beta^2L^2(t)+\beta^2 K(t)\big\}=0.
\end{eqnarray}
Then it is obvious  that (\ref{thm1-a})   is  a solution to (\ref{a7})  if $K(t), L(t), J(t)$  are solutions  to the differential equations  (\ref{thm1-b}), (\ref{thm1-c})  and (\ref{thm1-d}), respectively.

\hfill\fbox\\

Then we are going to solve the differential equations (\ref{thm1-b}), (\ref{thm1-c})  and (\ref{thm1-d}), respectively.

Let $$B:=2\beta^2, C:=2\alpha, D:=-\frac{1}{2(\sigma^2_f+\sigma^2_Q)},$$
then the Riccati  equation (\ref{thm1-b})  becomes
\begin{gather}\label{a9}
K^{'}(t)+BK^2(t)+CK(t)+D=0,\  \   K(T)=0.
\end{gather}
If $B\neq0,$ i.e. $\beta\neq 0,$ integrating
$$\frac{dK(t)}{BK^2(t)+CK(t)+D}=-dt$$
on  both sides with respect to $t$ we obtain that
\begin{gather}\label{a10}
\int \frac{dK(t)}{BK^2(t)+CK(t)+D}=-t+E,
\end{gather}
where $E$  is a constant. Since $\Delta=C^2-4BD=4\alpha^2+\frac{4\beta^2}{\sigma^2_f+\sigma^2_Q}>0,$  the   quadratic equation  $BK^2(t)+CK(t)+D=0$   has two different real roots given by
\begin{gather}\label{a8}
K_1, K_2=\frac{-C\pm \sqrt{C^2-4BD}}{2B}.
\end{gather}
Substituting (\ref{a8}) into (\ref{a10}) and considering the boundary condition $K(T)=0$   then we obtain
\begin{gather}\label{a13}
K(t)=\frac{K_1-K_2e^{B(K_1-K_2)(t-T)}}{1-(K_1/K_2)e^{B(K_1-K_2)(t-T)}}.
\end{gather}

If $B=0,$  i.e. $\beta=0,$ then
\begin{gather}\label{a14}
K(t)=\frac{1}{4\alpha(\sigma^2_f+\sigma^2_Q)}-\frac{1}{4\alpha(\sigma^2_f+\sigma^2_Q)}e^{2\alpha(T-t)}.
\end{gather}

With the value of  $K(t)$  defined in (\ref{a13}) or (\ref{a14}), the linear ordinary equation (\ref{thm1-c})  has the solution of the form
\begin{gather}\label{a15}
L(t)=e^{\int^T_t (\alpha+2\beta^2 K(s))ds}\big[\int^T_t -\frac{A_1}{\sigma^2_f+\sigma^2_Q}e^{\int^T_t -(\alpha+2\beta^2 K(y))dy} ds  \big].
\end{gather}

And the solution of (\ref{thm1-d})   is given by
\begin{eqnarray}\label{a16}
J(t)&=&\frac{u\theta}{r_d}(1-e^{r_d(T-t)})-\frac{\theta^2\sigma^2}{4r_d}(1-e^{2r_d(T-t)})
-\frac{A^2_1}{2(\sigma^2_f+\sigma^2_Q)}(T-t)\nonumber\\
&+&\int^T_t (\frac{1}{2}\beta^2L^2(s)+\beta^2K(s))ds.
\end{eqnarray}

From  \cite{FR93} the following verification theorem exists.
\begin{thm}
Let $W\in C^{1,2}([0,T]\times R^2)$  be a classical solution  to the HJB equation (\ref{3}) with the boundary condition $W(T,x,m)=u(x),$  then the value function $V$ given by (\ref{a0}) coincides with $W$  such that
$$W(t,x,m)=V(t,x,m).$$

In addition, let $\pi^{*}$  be the optimizer of (\ref{3}), that is for any $(t,x,m)\in [0,T]\times R^2$
$$V_t+\big[\pi^{*}(A_1+m)+xr_d+u   \big]V_x+\frac{1}{2}\big[ \sigma^2+\pi^{*2}(\sigma^2_f+\sigma^2_Q)  \big]V_{xx}
+\alpha m V_m+\frac{1}{2}\beta^2 V_{mm}=0.$$
Then $\pi^{*}(t, X^{*}_t, m(t))$ is the optimal strategy with
$$V_{\pi^{*}}(t,x,m)=V(t,x,m),$$
where $X^{*}_t$   is the surplus process under the optimal strategy $\pi^{*}.$
\end{thm}

From the above statements  we have the following results.
\begin{thm}\label{thm2}
With  the utility function (\ref{2}), the optimal strategy for the optimization problem (\ref{a0}) subject to (\ref{a3})  is
$$\pi^{*}_t=\frac{A_1+m(t)}{\theta(\sigma^2_f+\sigma^2_Q)}e^{-r_d(T-t)},\  \   \forall t\in [0, T].$$
And the value function is given by  the form
$$V(t,x,m)=\lambda-\frac{\gamma}{\theta}exp\big\{-\theta x e^{r_d(T-t)}+h(t,m)   \big\}$$
with $h(t,m)=K(t)m^2+L(t)m+J(t),$  where $K(t), L(t)$  and $J(t)$  are given by (\ref{a13})-(\ref{a16}).
\end{thm}

If $m(t)=0$  in (\ref{a2})  then  the   exchange rate price $Q(t)$ is degenerated into the process which is   modeled by means of geometric
Brownian motion
$$dQ(t)=Q(t)(u_Qdt +\sigma_Q dW^2_t).$$
In  this case, under the control of $\pi,$ $X^{\pi}_t$  satisfies the following stochastic equation
\begin{eqnarray}\label{b1}
d X_{t}^{\pi}=\big\{ \pi_t A_1+u+r_d X_t \big\}dt+\sigma dW_t+\pi_t \sigma_{f}d W^{1}_t+\pi_t \sigma_{Q}d W^{2}_t.
\end{eqnarray}
Then the HJB equation in (\ref{3})  becomes to be
\begin{gather}\label{b4}
V_t+\sup_{\pi}\Big\{\big[\pi A_1+xr_d+u\big] V_x+\frac{1}{2}\big[\sigma^2+\pi^2 (\sigma^2_f+\sigma^2_Q)\big]V_{xx}     \Big\}=0.
\end{gather}

By  solving the above HJB equation (\ref{b4})    the following corollary   is obtained.

\begin{cor}\label{cor1}
With  the $X^{\pi}_t$ in  (\ref{b1}) the optimal investment strategy is given by
$$\pi^{*}_t=\frac{A_1}{\theta(\sigma^2_f+\sigma^2_Q)}e^{-r_d(T-t)},\  \   \forall t\in [0, T].$$
Furthermore, the value function has the form
$$V(t,x)=\lambda-\frac{\gamma}{\theta}exp\big\{-\theta x e^{r_d(T-t)}+f(T-t)   \big\}$$
where
$$f(T-t)=\frac{\theta u}{r_d}(1-e^{r_d(T-t)})-\frac{\theta^2\sigma^2}{4r_d}(1-e^{2r_d(T-t)})-\frac{A^2_1}{2(\sigma^2_f+\sigma^2_Q)}(T-t).$$
\end{cor}


Let  $S^d_t$  be  the price of domestic risky asset described by the following stochastic differential equation
\begin{gather}
dS^d_t=S^d_t(u_d dt +\sigma_d dW^4_t),
\end{gather}
where  $u_d$  and $\sigma_d$  are positive constants  and $W^4_t$  is  the standard Brownian motion which is independent of $W_t.$
   Assume   the  insure   invests his   wealth   only in domestic currency market, i.e.,   domestic risk-free asset and domestic risky asset, then the surplus under the control $\pi$  is  that
\begin{gather}\label{b3}
dX^{\pi}_t=\big\{\pi (u_d-r_d)+u+r_d X_t   \big\} dt+\sigma dW_t+\pi \sigma_d d W^4_t.
\end{gather}

\begin{cor}\label{cor2}
In the domestic currency market,
the optimal strategy for the optimization problem (\ref{a0})   is
$$\pi^{*}_t=\frac{u_d-r_d}{\theta \sigma^2_d}e^{-r_d(T-t)},\  \   \forall t\in [0,T],$$
 and the  corresponding value function has the form
$$V(t,x)=\lambda-\frac{\gamma}{\theta}exp\big\{-\theta x e^{r_d(T-t)}+g(T-t)   \big\}$$
where
$$g(T-t)=\frac{\theta u}{r_d}(1-e^{r_d(T-t)})-\frac{\theta^2\sigma^2}{4r_d}(1-e^{2r_d(T-t)})-\frac{(u_d-r_d)^2}{2\sigma^2_d}(T-t).$$
\end{cor}
\begin{rem}\label{rem1}
By  comparing  the optimal investment  strategies and value functions in Corollary \ref{cor1}  and  Corollary \ref{cor2}, it is not  difficult to  see that \\
\hspace*{0.5cm} (i) Suppose  the insurer invests the same amount of his wealth into domestic risky assets  and foreign risky assets. We  can see  that

If $u_f+u_Q\ge u_d,$  then   the value function with exchange rate is always larger than the value function without exchange rate.
Conversely,  if  $u_f+u_Q<u_d$  it  is better for the insurer to invest in  domestic risky assets. \\
\hspace*{0.5cm} (ii)  Assume  that   the insurer wants to get the same value functions in the two kinds of currency markets.

  If $u_f+u_Q\ge u_d,$  in order to get the same value functions in the two cases, the amount of wealth invested in foreign risky assets are less  than that invested in domestic risky assets.
In addition,
 if $u_f+u_Q<u_d,$  in order to get the same values  of the value functions the insurer should invest more in foreign risky assets. \\
\hspace*{0.5cm} (iii)    When   $u_f+u_Q\ge u_d$  if the insurer invests more in foreign risky assets  the value function is higher than the value function in domestic risky assets.

\end{rem}

\begin{rem}
From  Corollaries  \ref{cor1} and  \ref{cor2},
it is not difficult to  see that  if no domestic risk-free assets are traded  and only   risky asset is considered even in domestic or foreign market, the optimal strategies are always constants, regardless of the level of wealth the insurer has.

\end{rem}

\section{Numerical examples and analysis }

In order to demonstrate our results,  numerical examples are presented for the optimal investment strategies and value functions in two kinds of  currency markets. Our objective is to study the effect of exchange rate on the insurer's decision and the value function. The particular numbers of  basic parameters are given in the following tables.

\begin{table}[h]
\centering
\begin{tabular}{|c|c|c|c|c|c|c|c|c|c|c|c|c|c|c|}

\hline
$T$ & $r_{d}$  & $\lambda$ & $\theta$ &$\gamma$ &  $u$& $\sigma$& $u_{f}$& $\sigma_{f}$& $u_{Q}$& $\sigma_{Q}$& $x$& $u_d$&$\sigma_d$ \\
\hline
4 & 0.1 &  1 & 1&1&0.4&0.1&0.3&$\sqrt{0.1}$&0.2&$\sqrt{0.3}$&2&0.3&$\sqrt{0.2}$ \\
\hline
\end{tabular}
\caption{}\label{Table 1}
\end{table}

\begin{figure}[h]
  \centering
  \subfigure[]
  {
  \begin{minipage}{7cm}\label{Fig 1}
  \centering
  \includegraphics[width=7cm]{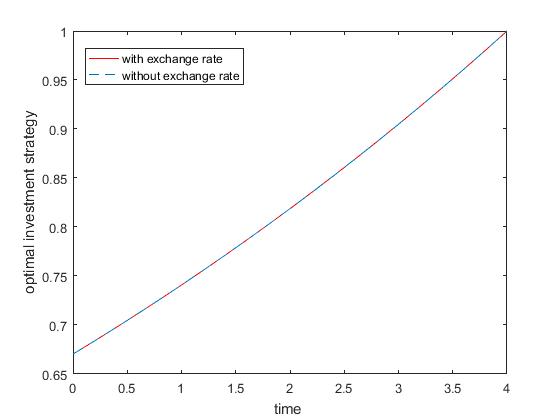}
  \end{minipage}
  }
 \subfigure[]
  {
  \begin{minipage}{7cm}\label{Fig 2}
  \centering
  \includegraphics[width=7cm]{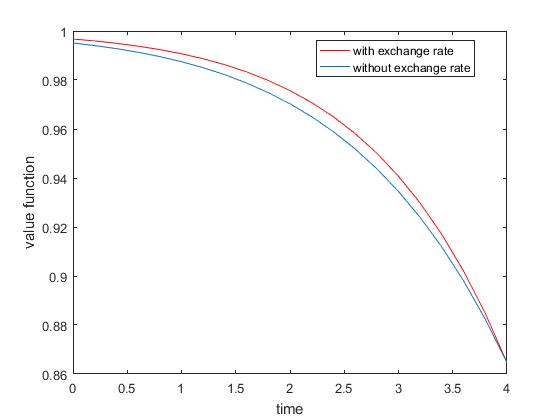}
  \end{minipage}
  }
\end{figure}

The numbers   in Table \ref{Table 1}  satisfy  that
$$\sigma_d^2(u_f+u_Q-r_d)=0.08=(\sigma^2_f+\sigma^2_Q)(u_d-r_d)$$  and
$$u_f+u_Q=0.5>u_d=0.3.$$
The graphs of the optimal investment strategies and value functions corresponding to the data in Table \ref{Table 1} are shown in (a) and (b).   They  show that if  the insure invests the same amount of its wealth into foreign and domestic risky assets, then the former produces a larger value function. Thus it is better for the insurer to invest in foreign risky assets. And  the  results in graphs (a)  and (b) also coincide with  the conclusions in (i)  of  Remark \ref{rem1}.

\begin{table}[h]
\centering
\begin{tabular}{|c|c|c|c|c|c|c|c|c|c|c|c|c|c|c|}

\hline
$T$ & $r_{d}$  & $\lambda$ & $\theta$ &$\gamma$ &  $u$& $\sigma$& $u_{f}$& $\sigma_{f}$& $u_{Q}$& $\sigma_{Q}$& $x$& $u_d$&$\sigma_d$ \\
\hline
4 & 0.1 &  1 & 1&1&0.4&0.1&0.3&0.2&0.2&$\sqrt{0.12}$&2&0.3&0.2 \\
\hline
\end{tabular}
\caption{}\label{Table 2}
\end{table}
\begin{figure}[h]
  \centering
  \subfigure[]
  {
  \begin{minipage}{7cm}\label{Fig 3}
  \centering
  \includegraphics[width=7cm]{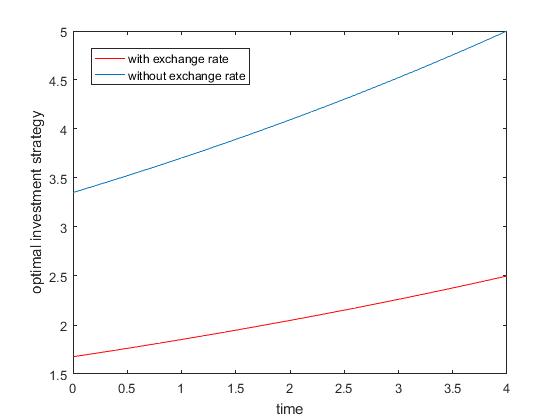}
  \end{minipage}
  }
 \subfigure[]
  {
  \begin{minipage}{7cm}\label{Fig 4}
  \centering
  \includegraphics[width=7cm]{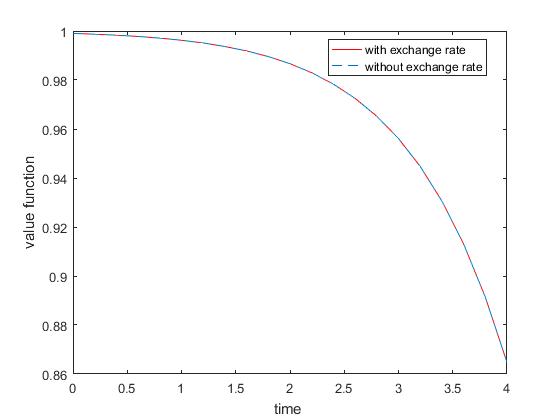}
  \end{minipage}
  }
\end{figure}

By employing   the numbers in Table \ref{Table 2}, the graphs of optimal investment strategies  and value functions are given in (c)  and (d). From  graphs (c)  and (d)  it is not difficult to see  that if the insurer wants to get the same value  functions in the currency markets  with and without exchange rates, he should invest more in domestic risky assets  than in foreign risky assets.
And the   numbers in Table \ref{Table 2}  satisfy  that
$$\sigma_d^2(u_f+u_Q-r_d)^2=(\sigma_f^2+\sigma_Q^2)(u_d-r_d)$$
and
$$u_f+u_Q>u_d.$$
Thus  graphs (c)  and (d) reflect the results in  (ii) of Remark \ref{rem1}.

The  graphs   (e)  and (f) are obtained from  the numbers listed  in Table \ref{Table 3}.
We first study the effect of the exchange rate on optimal investment strategies.  When the insurer has exponential preferences, the realization of their optimal investment strategies are illustrated in Figure (e). It can be seen that, when there are two currency markets the insurer invests a larger proportion of her wealth in the risky asset.

Secondly, we explore the effect of the exchange rate on the value function.
From figure (f),  we can  easily find that, it is much better for the insurer to invest her surplus in foreign risky asset  to decrease the risk. The value function with exchange rate is always larger than the value function without exchange rate, except the terminal value. It indicates that it is better to incorporate the exchange rate in the model.

\begin{table}[h]
\centering
\begin{tabular}{|c|c|c|c|c|c|c|c|c|c|c|c|c|c|c|}

\hline
$T$ & $r_{d}$  & $\lambda$ & $\theta$ &$\gamma$ &  $u$& $\sigma$& $u_{f}$& $\sigma_{f}$& $u_{Q}$& $\sigma_{Q}$& $x$& $u_d$&$\sigma_d$ \\
\hline
4 & 0.1 &  1 & 1&1&0.4&0.1&0.2&0.3&0.3&0.4&2&0.3&0.4 \\
\hline
\end{tabular}
\caption{}\label{Table 3}
\end{table}
\begin{figure}[h]
  \centering
  \subfigure[]
  {
  \begin{minipage}{7cm}\label{Fig 3}
  \centering
  \includegraphics[width=7cm]{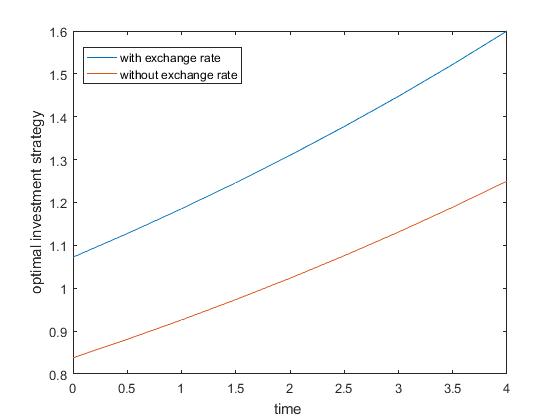}
  \end{minipage}
  }
 \subfigure[]
  {
  \begin{minipage}{7cm}\label{Fig 4}
  \centering
  \includegraphics[width=7cm]{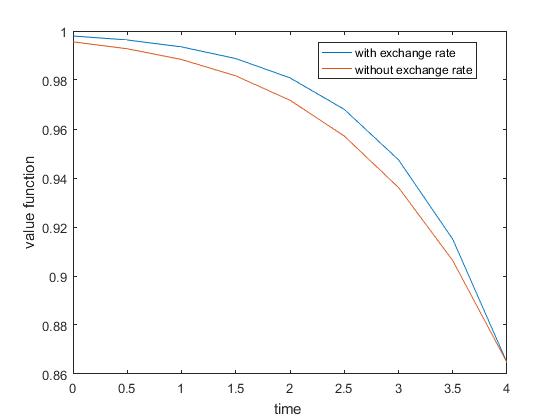}
  \end{minipage}
  }
\end{figure}

\bigskip


\end{document}